\def \be {\begin{equation}}
\def \eq {\end{equation}}
\def \bee {\begin{eqnarray}}
\def \eqq {\end{eqnarray}}

\def \bea {\begin{array}{c}}
\def \eqa {\end{array}}

\def \la {\left\langle}
\def \ra {\right\rangle}
\def \R {{\bf R}}

\def \Z {{\bf Z}}
\def \del {\partial}
\def \dels {\partial\kern-.5em / \kern.5em}
\def \As {{A\kern-.5em / \kern.5em}}
\def \Ds {D\kern-.7em / \kern.5em}

\def \a {\alpha}
\def \b {\beta}

\def \eps {\epsilon}

\def \lam {\lambda}

\def \II {I\hspace{-.1em}I\hspace{.1em}}

\def \IIB {\mbox{\II B\hspace{.2em}}}

\def \ys {{y\kern-.5em / \kern.3em}}
\def \O {{\cal O}}

\documentstyle[12pt]{article}

\def\[{\left [}
\def\]{\right ]}
\def\({\left (}
\def\){\right )}
\def\lbr{\left\{}
\def\rbr{\right\}}

\begin{document}
\begin{titlepage}
\begin{center}
\hfill   SISSA 58/98/FM  \\
\hfill   JHU-TIPAC-98007 \\
\hfill   hep-th/9806103 \\

\vskip .5in

{\large \bf D-Instanton in AdS$_5$ and Instanton in SYM$_4$}

\vskip .5in

Chong-Sun Chu$^{\dagger}$,
Pei-Ming Ho$^{\ddagger}$\footnote{on leave from Department of Physics,
University of Utah, Salt Lake City, UT 84112.},
Yi-Yen Wu$^{\S}$
\vskip .3in
{\em $^{\dagger}$ International School for Advanced Studies (SISSA)\\
Via Beirut 2, 34014 Trieste, Italy}
\vskip .3in
{\em $^{\ddagger}$ Department of Physics, Jadwin Hall, Princeton
University\\
Princeton, NJ 08544}
\vskip .3in
{\em $^{\S}$ Department of Physics and Astronomy, Johns Hopkins
University\\
Baltimore, Maryland 21218}

\end{center}

\vskip .5in

\begin{abstract}

Following the observation of Banks and Green that the D-instantons in AdS$_5$ correspond to the instantons in 4-dimensional supersymmetric Yang-Mills theory, we study in more detail this correspondence for individual  instantons. The supergravity solution for a D-instanton in AdS$_5$ is found using the ansatz used previously for D-instantons in flat space. We check that the actions and supersymmetries match between the D-instanton solution and the Yang-Mills instanton. Generalizing this result, we propose that any supergravity solution satisfying the ansatz corresponds to a (anti-)self-dual Yang-Mills configuration. Using this ansatz a family of
identities for correlation functions in the supersymmetric Yang-Mills theory are derived.

\end{abstract}
\end{titlepage}

\newpage
\renewcommand{\thepage}{\arabic{page}}
\setcounter{page}{1}
\setcounter{footnote}{0}

\section{Introduction}

Recently Maldacena \cite{Mald} has proposed a remarkable duality between certain d-dimensional superconformal field theory and string or M theory on a (d+1)-dimensional anti-de Sitter space (AdS$_{d+1}$) times a compact manifold. An important example of Maldacena's conjecture is that Type \IIB superstring theory on AdS$_5$$\times$S$^5$ is equivalent to the 
${\cal N}=4$ supersymmetric Yang-Mills (SYM) theory with gauge group $SU(N)$ in four dimensions. A precise recipe for relating Type \IIB supergravity (SUGRA) in the AdS bulk to the SYM on the AdS boundary has been given in \cite{Gubser,Witten1}. It was further explained by Witten \cite{Witten1} that this duality is actually {\it holographic} in the sense that Type \IIB superstring on AdS$_5$$\times$S$^5$ can be described by ${\cal N}=4$ SYM living on the boundary of AdS$_5$.

The holographic nature of Maldacena's conjecture is very interesting. According to 't Hooft \cite{Hooft}, a macroscopic region of quantum gravity can be described by a theory living on the boundary of this region. This holographic hypothesis was further developed by Susskind in
\cite{Susskind1} where string theory is argued to be a candidate for a holographic theory. Very recently holography in AdS space was studied in \cite{Susskind2,Barbon}. In this sense, one may regard Maldacena's conjecture as part of a more profound property of string theory:
holography. And, the many recent non-trivial tests of Maldacena's conjecture can be regarded as support to the holographic nature of string theory.

So far most of the tests of Maldacena's conjecture are about relating the perturbative particle-like modes of supergravity on AdS space to the operators of the boundary superconformal field theory \cite{Gubser,Witten1,test}; for example, Type \IIB supergravity on
AdS$_5$$\times$S$^5$ versus ${\cal N}=4$ SYM in four dimensions. On the other hand, one naturally expects more from Maldacena's conjecture. It is natural and perhaps more interesting to consider extended objects in string theory.  This question makes perfect sense in the sense of
holography, according to which one should be able to ask what the counterparts of these branes are in the boundary SYM, and  how they are manifested. At low energy, for this question one will consider Type \IIB SUGRA on a background which is AdS$_5$$\times$S$^5$ modified by the
presence of various branes compatible with this background. This SUGRA background will no longer be exactly an AdS$_5$$\times$S$^5$ space,\footnote{An exception is D-instanton. As we shall see, the SUGRA background which corresponds to adding D-instantons on
AdS$_5$$\times$S$^5$ is still AdS$_5$$\times$S$^5$ (in the Einstein frame).}  but will share the basic features of the AdS$_5$ space. Therefore, a very interesting subject associated with this question  is the study of supergravity solutions which correspond to adding various  kinds of
branes on AdS$_5$$\times$S$^5$.

Indeed, recently it was suggested in \cite{Banks1} and later in \cite{Witten2} that adding D-instantons to Type \IIB superstring on AdS$_5$$\times$S$^5$ should correspond to Yang-Mills (YM) instantons in the boundary SYM. Very recently various kinds of brane
configurations have been considered in AdS$_5$$\times$S$^5$ as well as its orientifold AdS$_5$$\times$RP$^5$ \cite{Witten2}, and they are argued to correspond to instantons, particles, strings, domain walls and baryon vertices in the boundary SYM. In principle, by generalizing the recipe given in \cite{Gubser,Witten1} to the cases with nontrivial backgrounds,
these AdS/CFT correspondences can be checked by studying the SUGRA partition function evaluated at the SUGRA background corresponding to branes in AdS$_5$$\times$S$^5$, and the SYM partition function by including the counterparts of branes in SYM. Stringy corrections correspond to $1/N$ corrections in the SYM. It is a difficult problem to include non-trivial R-R backgrounds in the string worldsheet. See however \cite{tsy} for some interesting recent works in this direction. 

In this paper, we will consider the case of a D-instanton smeared  over the transverse S$^5$. 
We show in more detail how a D-instanton is translated into a YM instanton on the gauge theory side, in agreement with \cite{Banks1}, with the interesting feature that the size of the YM
instanton is translated into the distance of a D-instanton from the AdS$_5$ boundary. As expected, the D-instanton action agrees with the YM instanton action and their supersymmetry (SUSY) can be matched. In fact, we notice that the action and SUSY matching hold for a larger class of configurations. We propose the generalization of this matching of instantons to the correspondence between any SUGRA solution satisfying the ansatz (\ref{ansatz}) and any (anti-)self-dual gauge configuration in SYM.

This paper is organized as follows. We start with a description of  AdS$_5$ in Sec.\ref{AdS} and match the moduli spaces between a D-instanton and a Yang-Mills instanton in Sec.\ref{match}. Then we consider an ansatz which preserves half of the original 32 SUSYs in AdS$_5$$\times$S$^5$ in Sec.\ref{theansatz} and in Sec.\ref{solution} we give the explicit
SUGRA solution for a smeared D-instanton. Briefly reviewing the SUSY and action for a YM instanton in Sec.\ref{YM-inst}, we show how to match them to the D-instanton in Sec.\ref{AdSCFT}, where we propose an AdS/CFT correspondence for more general configurations than just instantons. We also comment on how to understand the AdS/CFT  
correspondence for branes in anti-de Sitter space within the framework of \cite{Witten1} at
the lowest level. On the other hand, while the D-instantons are singular solutions, regular 
solutions of the ansatz in Sec.\ref{theansatz} can lead to identities for correlation functions in CFT, which will be shown in Sec.\ref{corr}.

\section{Review of AdS$_5$$\times$S$^5$}
\label{AdS}

We begin with a description of the Euclidean version of AdS$_5$. Define
AdS$_5$ in terms of the coordinates in $\R^{1,5}$
\be
-\eta^{AB}X_A X_B=X_{-1}^2-X_0^2-X_1^2-X_2^2-X_3^2-X_4^2=R^2,
\eq
where $\eta^{AB}=$diag$(-1,1,1,1,1,1)$ and $X_A$, $A=-1,0,1,\cdots,4$, transform linearly under $SO(1,5)$. $R$ is the radius of AdS$_5$ \cite{Mald}. Let
\be \label{coord}
U=\frac{X_{-1}+X_4}{R^{2}}, \quad V=R^{2}(X_{-1}-X_4),
\quad y_i=\frac{X_i}{UR},
\eq
so that the metric is
\be  \label{metric}
ds^2=\eta^{AB}dX_A dX_B= R^{2}\(\frac{1}{U^2}dU^2+U^2\eta^{ij}dy_i
dy_j\),
\eq
where $\eta^{ij}=$diag$(1,1,1,1)$ and $i=0,1,2,3$. Our description of AdS$_5$ is in accordance with that of \cite{Mald}, where $U$ has dimension mass, and $y_i$ has dimension 1/mass. Note that, with this choice of coordinates, the metric (\ref{metric}) contains an overall factor, $R^2$,
and therefore is naturally measured in units of $R^2$.

The following metric
\be \label{metric1}
d\tilde{s}^2=R^{2}\(\frac{1}{U^2}dU^2+U^2\eta^{ij}dy_i dy_j
+d\Omega_{5}^{2}\)
\eq
represents AdS$_5$$\times$S$^5$, where $(U,y_0,y_1,y_2,y_3)$ are coordinates of AdS$_5$, and $d\Omega_{5}$ denotes the volume element of the five-sphere S$^5$. By applying the Freund-Rubin ansatz \cite{Freund} to Type \IIB SUGRA with the self-dual 5-form field strength $F_{0123U}=R^{4}U^{3}$, eq.(\ref{metric1}) is a solution to the \IIB equations of motion
\cite{Schwarz}, $\,R_{\mu\nu}=\frac{1}{6}F_{\lambda\tau\rho\sigma\mu}
F^{\lambda\tau\rho\sigma}_{\hspace{0.7cm}\nu}$, and it represents the spontaneous compactification of Type \IIB theory on AdS$_5$$\times$S$^5$.

In the original argument of \cite{Mald} for the AdS/CFT duality based on considerations of D3-branes, the AdS$_5$$\times$S$^5$ space is the near horizon region of $N$ D3-branes with $R^2=\sqrt{4\pi gN}\alpha'$. In the low energy limit $\alpha'\rightarrow 0$, supergravity can be used as a good approximation of string theory if the curvature is small compared  with the
string scale, that is, $R^2\gg\alpha'$, or equivalently, $gN\gg 1$. We will take $gN$ large but fixed in the low energy limit. The D3-brane world volume corresponds to the boundary of AdS$_5$ at $U=\infty$.\footnote{The D3-brane metric is 
$ds^2=f^{1/2}(dr^2+r^2d\Omega_5^2)+f^{-1/2}dy^2$, where $f=1+R^4/r^4$. The asymptotically
flat region $r\gg R$ and the AdS$_5$ region $r\ll R$ are patched at  $r\sim R$. Here $r$ is related to $U$ by $U=r/R^2$. Therefore the D3-brane is located at $U\sim 1/R$ which goes to infinity as $\alpha'\rightarrow 0$.} ($U=0$ corresponds to the other end of the infinite throat within the horizon.)

The Killing spinors of AdS$_5$ which satisfy
$\(\del_{\mu}+\frac{1}{4}\omega^{ab}_{\mu}\gamma_{ab}\)\eps=0$ were
found in \cite{Lu}:
\be \label{K-spinor}
\eps=U^{1/2}\eps_0^+, \quad
\eps= \(U^{-1/2}+U^{1/2}\ys\)\eps_0^-,
\eq
where $\eps_0^{\pm}$ are constant spinors satisfying
$\gamma_U\eps_0^{\pm}=\pm\eps_0^{\pm}$.

\section{Matching D-Instanton and YM Instanton}
\label{match}

Recently it has been suggested \cite{Banks1,Witten2} that D-instantons of  Type \IIB superstring on AdS$_5$$\times$S$^5$ should correspond to YM instantons in the SYM living on the boundary of AdS$_5$ at $U=\infty$. In this paper we will try to study this correspondence. Let's begin with a D-instanton on AdS$_5$$\times$S$^5$. The general SUGRA  solution of a
D-instanton carrying R-R charge $Q^{(-1)}$ can naturally be decomposed\footnote{The meaning of this linear decomposition will become evident when we discuss eq.(\ref{eom5}).} into a zero mode (i.e., a constant) over S$^5$ and higher harmonics on S$^5$. The zero mode represents a  
D-instanton smeared over S$^5$ and the higher harmonics on S$^5$  do not carry any R-R
charge. Therefore, the general supergravity description of a D-instanton carrying charge $Q^{(-1)}$ can be given in terms of a smeared  D-instanton carrying the same charge $Q^{(-1)}$, 
and higher harmonics on S$^5$ which may be thought of as fluctuations around the smeared D-instanton.  As for AdS/CFT correspondence, it is then natural to suggest  that a D-instanton smeared over S$^5$ corresponds to a YM instanton in the SYM with maximal symmetry. More precisely, a D-instanton smeared over S$^5$ is invariant under the $SO(6)$ symmetry of S$^5$. Therefore, the corresponding YM instanton in ${\cal N}=4$ SYM should have $SO(6)$ symmetry;  that is, the six adjoint Higgs fields vanish.

Consider a D-instanton in AdS$_5$$\times$S$^5$ which is smeared  over S$^5$. It is a point in AdS$_5$, and its position in AdS$_5$ (i.e., moduli space) is
$(U,y_0,y_1,y_2,y_3)$. On the other hand, a YM instanton which has  $SO(6)$ symmetry in the boundary SYM is described by its position at the AdS$_5$ boundary and its size $L$, and therefore its moduli space is $(L,x_0,x_1,x_2,x_3)$. Both $(U,y_0,y_1,y_2,y_3)$ and  
$(L,x_0,x_1,x_2,x_3)$ transform under $SO(1,5)$ (as isometry and conformal symmetry
respectively), and therefore the two moduli spaces can be identified. First we can identify
$(y_0,y_1,y_2,y_3)$ with the coordinates $(x_0,x_1,x_2,x_3)$ of the  AdS$_5$ boundary. To see how $U$ is related to $L$, we consider an $SO(1,5)$ isometry of AdS$_5$ which acts on a D-instanton's position as
\be \label{scaling}
(U,y_0,y_1,y_2,y_3) \rightarrow
(\lam U,\lam^{-1}y_0,\lam^{-1}y_1,\lam^{-1}y_2,\lam^{-1}y_3),
\quad V\rightarrow \lam^{-1}V,
\quad X_{0,1,2,3}\rightarrow X_{0,1,2,3}.
\eq
Eq.(\ref{scaling}) should induce an $SO(1,5)$ conformal transformation on the AdS$_5$ boundary at $U=\infty$ \cite{Witten1} which acts on a YM instanton's size and position, $(L,x_0,x_1,x_2,x_3)$, as follows.
\be \label{scaling1}
(L,x_0,x_1,x_2,x_3) \rightarrow
(\lam^{-1}L,\lam^{-1}x_0,\lam^{-1}x_1,\lam^{-1}x_2,\lam^{-1}x_3).
\eq
Eq.(\ref{scaling}) is consistent with (\ref{scaling1}) if $L=1/U$  (up to a constant coefficient). Therefore, a D-instanton positioned at $U$ of AdS$_5$ corresponds to a YM instanton of size $1/U$ in the boundary SYM. (See also (\ref{size}) below for an explicit derivation of this  fact using the D-instanton solution presented later). As a D-instanton  approaches the AdS$_5$ boundary (i.e., $U\rightarrow\infty$), in the SYM it corresponds to a YM instanton in the zero-size limit ( i.e., $L\rightarrow 0$).  Since the AdS$_5$ boundary corresponds to the D3-brane world volume in the original argument for the AdS/CFT duality, this correspondence is consistent with the well-known fact that a D$(p-4)$-brane localized within D$p$-branes is an instanton in the zero-size limit in the D$p$-brane world volume theory \cite{ton}.

The observation of $L=1/U$ is very natural from the point of view of holography: a length scale in the boundary CFT gets transformed into a perpendicular coordinate in the AdS bulk. Obviously this is a general holographic feature of anti-de Sitter space. Very recently a similar
observation has been  made in \cite{Susskind2}. Holography in AdS can be pictured vividly according to \cite{Susskind2}: The scale size  ($L$) of an object in the boundary theory gets transmuted into a spatial dimension ($U$). Furthermore, $L=1/U$ is consistent with the I.R.-U.V. connection proposed in \cite{Susskind2}, which says that being small in size in the SYM corresponds to being far away from the center (close to the boundary) of anti-de Sitter space.

\section{The Ansatz for D-instanton} \label{theansatz}

The supergravity solution for D-instantons of Type \IIB superstring theory on flat space has been constructed in \cite{Gibbons} where the metric $g_{\mu\nu}$, the dilaton $\phi$ and the R-R axion $\chi$ are the only non-vanishing fields. Here we will construct the supergravity solution for D-instantons on AdS$_5$$\times$S$^5$. {\it A priori} the non-vanishing fields involved are $g_{\mu\nu}$,  $\phi$, $\chi$ and the self-dual 5-form field strength
$F_{\mu\nu\rho\lambda\sigma}$, where $F_{\mu\nu\rho\lambda\sigma}$ is
expected to trigger the spontaneous compactification of Type \IIB on AdS$_5$$\times$S$^5$ 
\cite{Freund,Schwarz}. The bosonic part of  Type \IIB supergravity action in the Einstein frame is
\be \label{SUGRAaction}
I_{SUGRA}=\frac{1}{2\kappa_{10}^2}\int d^{10}x \lbr\sqrt{-g\,}\left[
R-\frac{1}{2}(\del\phi)^2 -\frac{1}{2}e^{2\phi}(\del\chi)^2 \right]
-\del_{\mu}\left(e^{2\phi}\chi g^{\mu\nu}\del_{\nu}\chi\right)\rbr
+ \cdots,
\eq
where $\kappa_{10}=8\pi^{7/2}g\alpha'^2$ and the last term is a  surface term which is given by the total flux of the 9-form field strength dual to $\chi$ \cite{Green}. The part of self-dual gauge field $F_{\mu\nu\rho\lambda\sigma}$ cannot be written down in (\ref{SUGRAaction}), but will be included in the equations of motion. We shall begin with the relevant Type \IIB supergravity equations of motion in the Einstein frame \cite{Schwarz}. For D-instantons, it is appropriate to do Wick rotation and $\chi\rightarrow i\chi$ so that we have
\be \label{eom1}
R_{\mu\nu} - \frac{1}{2}(\del_{\mu}\phi)(\del_{\nu}\phi)
+ \frac{1}{2}e^{2\phi}(\del_{\mu}\chi)(\del_{\nu}\chi) =
\frac{1}{6}F_{ \lambda_{1}\lambda_{2}\lambda_{3}\lambda_{4}\mu}
F^{\lambda_{1}\lambda_{2}\lambda_{3}\lambda_{4}}_{\hspace{1.4cm}\nu},
\eq
\be \label{eom2}
\nabla_{\mu}\( e^{2\phi}\del^{\mu}\chi \) = 0,
\eq
\be \label{eom3}
\nabla^{\mu}\nabla_{\mu}\phi +
e^{2\phi}(\del^{\mu}\chi)(\del_{\mu}\chi) = 0,
\eq
and the 5-form field strength $F_{\mu\nu\rho\lambda\sigma}$ satisfies self-duality condition.

For the solution of a D-instanton which preserves half of the supersymmetries, we take the following ansatz
\be \label{ansatz}
\chi - \chi_{\infty} \,=\, \pm\(e^{-\phi}-e^{-\phi_{\infty}}\),
\eq
where the constants $\phi_{\infty}$ and $\chi_{\infty}$ are the values of the fields $\phi$ and $\chi$ at the AdS$_5$ boundary.

Now we justify this ansatz (\ref{ansatz}) by showing that it leads to a preservation of half of the SUSYs. What we do here parallels the discussion on D-instantons in a flat background in \cite{Gibbons}. Since all the fermionic fields vanish, the only SUSY transformations we need to check are \cite{Schwarz}
\bee
\delta\lambda&=&-\frac{1}{\tau_2}\left(\frac{\tau^*-i}{\tau+i}\right)
\gamma^{\mu}\(\del_{\mu}\tau\)\eps^*, \\
\delta\psi_{\mu}&=&\left(D_{\mu}-\frac{i}{2}Q_{\mu}\right)\eps,
\label{psi}
\eqq
where
\bee
D_{\mu}&=&\del_{\mu}+\frac{1}{4}\omega^{ab}_{\mu}\gamma_{ab}, \\
Q_{\mu}&=&-\frac{1}{4\tau_2}\left\{\left(\frac{\tau-i}{\tau^*-i}\right)
\del_{\mu}\tau^*+\left(\frac{\tau^*+i}{\tau+i}\right)
\del_{\mu}\tau\right\},
\eqq
and we denote $\tau=\tau_1+i\tau_2=\chi+i e^{-\phi}$. The $\lambda$ and $\psi$ are the complex spin $1/2$ and $3/2$ fields in \IIB supergravity, respectively, and the SUSY parameter
$\eps=\eps_1+i\eps_2$ is a complex spinor in ten dimensions. In order to study the SUSY transformations in Euclidean space, we need to replace the complex number $i$ by another algebraic element $e$ which satisfies $e^2=1$ and $e^*=-e$ \cite{Gibbons}. The ansatz (\ref{ansatz}) becomes $d\tau=(1\pm e)d\tau_1$ and thus $\delta\lambda=0$ is satisfied for $\eps_1=\pm\eps_2$. It also follows from the ansatz that
\be
Q_{\mu}=\mp \frac{1}{2}\del_{\mu}
\log \left( \frac{\tau_2}{\tau_2+a}
\right),
\eq
where $a=\(1 -e^{-\phi_{\infty}} \pm \chi_{\infty}\)/2$.
Thus (\ref{psi}) becomes
\be
\delta\psi_{\mu}=(1\pm e)h D_{\mu}\(h^{-1}\eps_1\),
\eq
where $h=\(1+a/\tau_2\)^{1/4}$. Given the Killing spinors $\eps'$ on AdS$_5$$\times$S$^5$, i.e., $D_{\mu}\eps'=0$, the SUSYs preserved by the instanton background are $\eps_1=(1\pm e)h\eps'$. Half of the SUSYs are preserved.

We will give the SUGRA solution for smeared D-instantons on AdS$_5$ in the following section. 
The choice of sign in (\ref{ansatz}) determines whether the solution eqs.(\ref{sol}) and (\ref{sol1})
corresponds to a D-instanton or an anti-D-instanton (see eq.(\ref{charge1}) below). With this ansatz the contribution of $\phi$ and $\chi$ to the energy-momentum tensor in (\ref{eom1}) vanishes, and therefore eqs.(\ref{eom1})-(\ref{eom3}) are reduced to the following:
\be \label{eom4}
R_{\mu\nu} =
\frac{1}{6}F_{ \lambda_{1}\lambda_{2}\lambda_{3}\lambda_{4}\mu}
F^{\lambda_{1}\lambda_{2}\lambda_{3}\lambda_{4}}_{\hspace{1.4cm}\nu},
\eq
\be \label{eom5}
\frac{1}{\sqrt{g}}\del_{\mu}\(g^{\mu\nu}\sqrt{g}\del_{\nu}e^{\phi}\)=0.
\eq
As discussed in Sec.2, it is clear that eq.(\ref{eom4}) is exactly solved by the AdS$_5$$\times$S$^5$ solution (as defined in (\ref{metric1})) by applying the Freund-Rubin ansatz \cite{Freund}. Hence the SUGRA background which corresponds to adding
D-instantons to Type \IIB superstring on AdS$_5$$\times$S$^5$ is still AdS$_5$$\times$S$^5$ (in the Einstein frame). Note that, because eq.(\ref{eom5}) is linear in $e^{\phi}$, the most
general solution of $e^{\phi}$ on AdS$_5$$\times$S$^5$ is naturally a sum of spherical harmonics of S$^5$. According to ansatz (\ref{ansatz}), $\chi$ is determined by $e^{-\phi}$. In the following discussion, the coordinates and metric of AdS$_5$$\times$S$^5$ defined in (\ref{metric1}) will be employed explicitly.

\section{SUGRA Solution of D-Instanton in AdS$_5$} \label{solution}

Eq.(\ref{eom5}) is just  the wave equation in AdS$_5$$\times$S$^5$ for a massless scalar
field and so it can be solved easily.  As discussed in Sec.3, the D-instanton smeared over S$^5$ is to be considered to match the YM instanton in SYM. In the following, we consider a smeared D-instanton localized at a point $(U_0,y^0_0,y^1_0,y^2_0,y^3_0)$ of AdS$_5$. Eq.(\ref{eom5}) can be solved in terms of the $SO(1,5)$-invariant variable
\be
d^2=\frac{\eta^{AB}X_A X_B}{R^2}=
U_0 U\[ (\frac{1}{U}-\frac{1}{U_0})^2 + \sum_{i=0}^{3}(y^i-y^i_0)^2 \].
\eq
Note that $d^2$ is dimensionless (measured in units of $R^2$), and it has no explicit dependence on $R$. There is thus no explicit dependence on $R$ at all in $\phi$ and $\chi$. The solution for eq.(\ref{eom5}) is
\be \label{xyz}
e^{\phi}=c_0+c_1\frac{(d^2+2)(d^4+4d^2-2)}{d^{3}\cdot(d^2+4)^{3/2}},
\eq
where both $c_0$ and $c_1$ are dimensionless constants.

Notice that this solution is more general than the solution (2.19) in \cite{Witten1}. There, the solution has a source localized on the AdS boundary. Our solution here has a source term localized at an arbitrary point in AdS$_5$. To recover their solution from ours, we have to take the $U_0 \rightarrow \infty$ limit of (\ref{xyz}) with $c_0 =0$, $c_1\sim U_0^4$, and use the large $d^2$ expansion of this solution
\be
e^{\phi}=e^{\phi_{\infty}}\;-\;\frac{6 c_1}{d^8}+\cdots,
\eq
where $e^{\phi_{\infty}}=c_{0}+c_{1}$.

For large $U$ and $y^i \neq y^i_0$, $1/U_0 \neq 0$, the solutions of $e^{\phi}$ and $\chi$ are
\be \label{sol}
e^{\phi}=e^{\phi_{\infty}}\;-\;\frac{6 c_1}{U^4}Z^4(y;y_0,U_0),
\eq
\be \label{sol1}
\chi=\chi_\infty \;\pm\; \frac{6
c_1\!\cdot\!e^{-2\phi_{\infty}}}{U^4}Z^4(y;y_0,U_0),
\eq
where the choice of sign in (\ref{sol1}) is in accordance with (\ref{ansatz}), and
\be \label{S}
Z(y;y_0,U_0)=\frac{1}{U_{0}\[\sum_{i=0}^{3}(y^i-y^i_0)^2
\, + \,\frac{1}{U_{0}^2}\]}.
\eq
The $1/U^4$ dependence in the second term in (\ref{sol})-(\ref{sol1}) is expected in general for a massless scalar field in AdS$_5$. Note that $Z(y;y_0,U_0)$ is a bell-shaped function centered at $(y^0_0,y^1_0,y^2_0,y^3_0)$ of size $1/U_{0}$. It is the profile of the source (YM instanton) in CFT corresponding to the D-instanton in AdS. In the limit $U_0\rightarrow\infty$, we have
\be
Z^4\;\;\rightarrow\;\;  \frac{\pi^2}{6}\prod_{i=0}^{3}\delta(y^i-y^i_0),
\eq
which corresponds to a point-like YM instanton at $(y^0_0,y^1_0,y^2_0,y^3_0)$. In another limit where $U_0\rightarrow 0$, $Z^4$ vanishes. (To be more precise, $Z^4$ is a delta function at
$U_0= \infty$.)

It is well known that D-instanton and D7-brane carry electric and  magnetic R-R charges respectively, and the charges satisfy Dirac quantization condition \cite{Dirac}. The constant $c_1$ in our D-instanton  solution can be quantized this way. Note that the D-instanton under consideration here is a D-instanton smeared over S$^5$. It can be regarded either as a
five-sphere in AdS$_5$$\times$S$^5$ or as a point (D-instanton) in  AdS$_5$. The corresponding magnetic object should be a D7-brane wrapped on S$^5$. The nine-form field strength $F^{(9)}$ dual to the R-R one-form field strength of a D-instanton smeared over S$^5$ is
$F^{(9)}=\epsilon^{(5)}\cdot F^{(4)}$, where $\epsilon^{(5)}$ is the five-index $\epsilon$ tensor of S$^5$, and the four-form field strength $F^{(4)}$ depends only on AdS$_5$. One can also think of a smeared D-instanton and a wrapped D7-brane in AdS$_5$$\times$S$^5$ as a D-instanton and a D2-brane in AdS$_5$. Then $F^{(4)}$ is carried by the D2-brane and is the dual of R-R one-form field strength in AdS$_5$.

We begin with the D-instanton charge. The Noether current $J_{\mu}$ associated with the shift symmetry of R-R axion $\chi$ ($\chi\,\rightarrow\,\chi +\mbox{constant}$) is $J_{\mu}=e^{2\phi}\del_{\mu}\chi$. Conservation of $J_{\mu}$ follows directly from 
eq.(\ref{eom2}).  The Noether charge (or D-instanton  charge) $Q^{(-1)}$ defines the R-R charge carried by our D-instanton solution, and is given by
\be \label{charge}
Q^{(-1)} = V_5 \cdot\int_{\del(\!AdS_{5}\!)} J_{\mu} d\Sigma^{\mu},
\eq
where $\del(\!\mbox{AdS}_{5}\!)$ denotes the AdS$_5$ boundary at $U=\infty$, and the integration is performed over $\del(\!\mbox{AdS}_{5}\!)$. $V_5 =  \pi^{3}R^{5}$ is the volume of the five-sphere and it appears as an overall factor in $Q^{(-1)}$ because D-instanton is smeared over  S$^5$. Therefore the charge $Q^{(-1)}$ carried by our D-instanton solution,
(\ref{xyz})-(\ref{sol1}), is
\be \label{charge1}
Q^{(-1)} = \mp 4\pi^{2}R^{3} V_5 \cdot c_{1}.
\eq
According to \cite{Gibbons,Green}, it is straightforward to calculate the Euclidean action of our D-instanton, $I_{D}$. It follows from (\ref{ansatz}) that the kinetic terms of $\phi$ and $\chi$ cancel each other in the action and the only contribution comes from the surface term in (\ref{SUGRAaction}).\footnote{This surface term in general has contributions from the AdS boundaries at both $U=\infty$ and  $U=0$.} It is
\be \label{action}
I_{D} = \frac{Q^{(-1)}}{2\kappa_{10}^{2}g},
\eq
where $g=e^{\phi_{\infty}}$ is the string coupling constant.

Together with the magnetic charge $P^{(2)}$ carried by a D7-brane wrapped on S$^5$, the electric charge $Q^{(-1)}$ of a smeared D-instanton satisfies the following Dirac quantization condition \cite{Dirac,Joe}.
\be \label{condition}
\(2\kappa_{10}^{2}\)^{-1}\cdot Q^{(-1)} P^{(2)} = 2 \pi n, \quad  n\in\Z.
\eq
Similar to \cite{Gibbons}, it can be shown that a D7-brane has a minimal quantized magnetic charge $P^{(2)}=1$ due to the $SL(2,\Z)$ transformation $\chi\rightarrow \chi+1$. Therefore $c_1$, $Q^{(-1)}$ and $I_{D}$ of our D-instanton solution are quantized as follows.\footnote{One
may also regard smeared D-instanton and wrapped D7-brane in AdS$_5$$\times$S$^5$ as
D-instanton and D2-brane in AdS$_5$, and apply Dirac quantization condition to D-instanton and D2-brane charges. The result of this quantization in AdS$_5$ is equivalent to (\ref{q1})-(\ref{q3}) by noting that the five-dimensional Newton constant, $\kappa_{5}$, of AdS$_5$ is related to $\kappa_{10}$ by $\kappa_{5}^2= \kappa_{10}^{2}/V_5 $.}
\be \label{q1}
c_1 = \frac{2\kappa_{10}^{2}}{2\pi R^{3} V_5}\cdot n
=\frac{4\pi n}{N^2},
\eq
\be \label{q2}
Q^{(-1)} = 2\kappa_{10}^{2}\cdot 2 \pi n,
\eq
\be \label{q3}
I_D = \frac{2 \pi n}{g}.
\eq
Since eq.(\ref{eom5}) is linear in $e^{\phi}$, the  superposition of $K$ D-instanton solutions is again a solution which also preserves half of the SUSYs and is interpreted as a configuration consisting of $K$ D-instantons. The charge $Q^{(-1)}$ and the action $I_D$ of this
configuration are simply
\be \label{charge2}
Q^{(-1)}=2\kappa_{10}^{2}\cdot\sum_{k=1}^{K} 2 \pi n^{(k)},
\eq
\be \label{ID}
I_D = \frac{\sum_{k=1}^{K} 2 \pi n^{(k)}}{g},
\eq
where $n^{(k)}$ defines the charge of the $k$-th D-instanton. This feature of superposition is closely related to the fact that D-instantons preserve half of the SUSYs, and therefore is not affected by quantum corrections.

\section{Instanton in ${\cal N}=4\,$ SYM Theory} \label{YM-inst}

The ${\cal N}=4$ $SU(N)$ SYM theory in $4$ dimensions can be obtained by dimensional reduction from the $10$ dimensional SYM theory. The action is
\be
I_{SYM}=\frac{1}{g_{YM}^2}\int d^4
y\,\mbox{Tr}\!\(\frac{1}{2}[X_{\mu}, X_{\nu}]^2
-\bar{\Psi}\Gamma^{\mu}[X_{\mu},\Psi]\),
\eq
where $\mu,\nu=0,1,\cdots,9$. $X_i=-iD_i$ ($i=0,1,2,3$) are covariant derivatives and $X_a$ ($a=4,\cdots,9$) are adjoint Higgs fields.  $X_{i}$ and  $X_a$ are hermitian.  $\Psi$ is a Majorana-Weyl spinor in 10 dimensions and it decomposes into four Majorana spinors in 4 dimensions.

The Poincar\'{e} supersymmetry transformation on the fields is
\bee
\delta X_{\mu}&=&i\bar{\xi}\Gamma_{\mu}\Psi, \label{dX1}\\
\delta\Psi&=&\frac{i}{2}[X_{\mu},X_{\nu}]\Gamma^{\mu\nu}\xi.
\label{dPsi1}
\eqq
This symmetry is enhanced into the ${\cal N}=4$ superconformal group by the following special supersymmetry transformations \cite{Mehta}
\bee
\delta X_{\mu}&=&i\bar{\zeta}\ys\Gamma_{\mu}\Psi, \label{dX2}\\
\delta \Psi&=&\frac{i}{2}[X_{\mu},X_{\nu}]\Gamma^{\mu\nu}\ys\!\zeta
+2 X_a\Gamma^a\zeta. \label{dPsi2}
\eqq

For a (anti-)self-dual gauge field configuration which satisfies $F_{ij}=\pm\frac{1}{2}\eps_{ijkl}F^{kl}$ with $X_a=0$ and $\Psi=0$,
half of the SUSYs are preserved. The preserved SUSYs are parametrized by the $\xi$ of positive chirality ($\Gamma^{0123}\xi=\xi$) and the $\zeta$ of negative chirality ($\Gamma^{0123}\zeta=-\zeta$). The broken SUSYs generate 16 fermionic zero modes for the (anti-)self-dual configuration. These correspond to the 16 fermionic zero modes for a D-instanton in AdS$_5$$\times$S$^5$.

The gauge field configuration for an $SU(2)$ instanton of size $L$ centered at $y_0$ is given by
\be \label{size}
F_{ij}=4 Z^2(y;y_0,1/L)\sigma_{ij},
\eq
where $Z(y;y_0,1/L)$ is defined by (\ref{S}) and $\sigma_{0a}=-\frac{1}{2}\sigma_a$,
$\sigma_{ab}=\frac{1}{4i}\[\sigma_a,\sigma_b\]$ for $a,b=1,2,3$. This demonstrates again that the AdS$_5$ coordinate
\be \label{LU}
U_0 =1/L
\eq
of a D-instanton is translated into the size of the YM instanton. It is amusing that the same function $Z$ used in the SUGRA solution also appears here. The action for $K$ YM instantons is
\be \label{S-I}
I_{SYM}=\frac{4\pi}{g_{YM}^2} \sum_{k=1}^{K} 2 \pi n^{(k)},
\eq
where $n^{(k)}$ is the charge of the k-th YM instanton.

\section{AdS/CFT Correspondence} \label{AdSCFT}

In this section we first consider the matching of supersymmetry between the SUGRA on AdS and SYM on the boundary. For the case with no instantons, the Killing spinors of AdS$_5$$\times$S$^5$ defined in (\ref{K-spinor}) behave at large $U$ as
\be \label{4}
\eps\sim U^{1/2}\eps_0^+, \quad \eps\sim U^{1/2}\ys\eps_0^-.
\eq
These expressions\footnote{For simplicity, we have not written  explicitly in (\ref{4})  the Killing spinor on the $S^5$ part. This part of the spinor will give rise to a {\bf 4} of $SU(4)$ which is essential for the matching here.} coincide with the SUSY parameters $\xi$ and  $\ys\!\zeta$
in ${\cal N}=4$ SYM (\ref{dX1})-(\ref{dPsi2}) up to a factor of  $U^{1/2}$. It is thus natural to identify that
\be
\left(\begin{array}{c}\xi\\ \zeta\end{array}\right)\sim
\left(\begin{array}{c}\eps_0^+\\ \eps_0^-\end{array}\right).
\eq

In the background of (\ref{ansatz}), only half of the SUSYs corresponding to
$\eps_1=\eps_2$ or $\eps_1=-\eps_2$ are preserved. On the other hand, for a (anti-)self-dual background field in SYM, the preserved SUSYs correspond to $\Gamma^{0123}\xi=\xi$ and $\Gamma^{0123}\zeta=-\zeta$. Let  $\xi_{\pm}$ be defined by $\Gamma^{0123}\xi_{\pm}=\pm\xi_{\pm}$ and similarly for $\zeta_{\pm}$. The correspondence should be
\be \label{spin-map}
\left(\begin{array}{c}\xi_{\pm}\\ \zeta_{\pm}\end{array}\right)
\sim\left(\begin{array}{c}(\eps_0^+)_r\pm(\eps_0^+)_i\\
(\eps_0^-)_r\mp(\eps_0^-)_i\end{array}\right),
\eq
where $(\eps_0^{\pm})_r$ and $(\eps_0^{\pm})_i$ are the real and imaginary parts of $(\eps_0^{\pm})$, respectively. Hence we have obtained the matching of SUSY parameters without explicitly comparing the SUSY algebras.

Note that most of the discussions in this paper apply to any SUGRA solution satisfying the ansatz (\ref{ansatz}) and any (anti-)self-dual gauge field background in SYM. We are thus led to  
propose a more general correspondence than the identification of instantons: We conjecture that {\em all solutions satisfying (\ref{ansatz}) in AdS$_5$$\times$S$^5$ correspond to (anti-)self-dual gauge field configurations in SYM}. Obviously the ansatz (\ref{ansatz}) admits many other solutions different from our solution for smeared D-instantons, including those with
non-trivial dependence on S$^5$. It would be very interesting to see what are their corresponding configurations in SYM.

The first check one can do for the conjecture is the action. The ansatz (\ref{ansatz}) implies that the SUGRA action is given solely by the surface term in (\ref{SUGRAaction}), which equals the instanton number up to a constant as given by (\ref{ID}). Correspondingly, for (anti-)self-dual
gauge configurations the Yang-Mills action equals the instanton number up to an overall factor as in (\ref{S-I}). Thus the two actions agree by identifying $g=g_{YM}^2/4\pi$.

According to the AdS/CFT correspondence, we have \cite{Gubser, Witten1}
\be \label{dual}
\la e^{\int d^4 y \psi_{\infty}\O}\ra_{CFT}=e^{-I_{SUGRA}(\psi)},
\eq
where $\psi$ represents supergravity fields at the AdS background with boundary values $\psi_{\infty}$. $I_{SUGRA}(\psi)$ is the SUGRA action for the configuration $\psi$. $\O$ stands for the operator in CFT which couples to $\psi_{\infty}$. For a scalar field $\psi$ of mass $m$, the field goes like $\psi_{\infty} U^{\lam_+}$ when $U\rightarrow\infty$, where $\lam_+$ is the
larger root of $\lam(\lam+4)=m^2$ \cite{Gubser,Witten1}. The explicit expression of the operator $\O$ can be derived by studying the Dirac-Born-Infeld action (with Wess-Zumino terms) for the D3-branes in the AdS background \cite{deAlwis,Das}.

In general, the asymptotic behavior of the massless field $e^{\phi}$ of (\ref{eom5}) in AdS$_5$ is of the form
\be \label{ppp}
e^{\phi} = g(y) + h(y)U^{-4}, \quad\quad \mbox{at large $U$.}
\eq
Thus the boundary value of $e^{\phi}$ that couples to an operator in SYM is $g(y)$. Solution with this asymptotic behavior (\ref{ppp}) can be regular or singular. We first consider the singular case. The regular case will be discussed in the next section. The D-instanton solution (\ref{xyz})
we have above is singular at the position of the D-instanton. Since the SUGRA solution of D-instanton modifies the dilaton and axion fields only by terms decaying like $1/U^4$ for large $U$, the D-instantons do not couple to any operator in SYM directly in the sense of (\ref{dual}).\footnote{At any rate, it is not even appropriate to use  a singular solution in (\ref{dual}) as it is important in the work of \cite{Witten1} that the field $\psi$ has to be regular everywhere on AdS so that $\psi$ is uniquely determined by its boundary value $\psi_{\infty}$.}
This distinction between singular SUGRA solutions as backgrounds and regular solutions as fluctuations is analogous to the distinction between non-normalizable and normalizable SUGRA modes in the Lorentzian formulation of the AdS/CFT correspondence
\cite{Bala}.

The above discussion suggests a possible way to study AdS/CFT correspondence in the presence of D-instantons (or other D-branes in general). What one can do is to consider regular SUGRA solutions in the presence of the D-instanton background and modify eq.(\ref{dual}).
According to the correspondence between the supergravity on AdS and  SYM on the boundary, the partition functions for the two theories  should agree when evaluated at corresponding backgrounds. Thus, in the presence of instantons,  (\ref{dual}) should be understood as follows,
\be \label{dual1}
\la e^{\int d^4 y \psi_{\infty}\O}\ra_{CFT}^{(b)}=
e^{-I_{SUGRA}(\psi;B)},
\eq
where $b$ represents an instanton background of SYM with a  corresponding D-instanton background $B$ of SUGRA on AdS. The field  $\psi$ is now understood as fluctuations above the background $B$,  and $\O$ represents the corresponding SYM fluctuations above the  
background $b$. The superscript $(b)$ in $\la\cdot\ra_{CFT}^{(b)}$  denotes that the correlation function is evaluated with the background $b$; and similarly $I_{SUGRA}(\psi;B)$ is the effective action for $\psi$ in the background of $B$.

For example, consider a scalar field $\psi$ of mass $m$ on AdS$_5$. At a nontrivial background its wave equation is typically modified to be
\be
\frac{1}{\sqrt{g}}\del_{\mu}\(g^{\mu\nu}\sqrt{g}\del_{\nu}\psi\)-m^2\psi-V'(\psi)=0,
\eq
where $V(\psi)$ is the effective potential induced by the background. The appearance of the potential $V$ will modify the contribution of  $\psi$ to supergravity action as well as the Green's function which  determines $\psi$ from its boundary values. The usual perturbation  theory can be used to solve for the modified Green's functions and  the supergravity action. Then (\ref{dual}) implies that the correlation functions in SYM for the operator coupled to  
$\psi_{\infty}$ should also be modified. This should be understood as the effect of a corresponding background in SYM. Therefore, in principle one can check \cite{chw} the correspondence between a SUGRA  background and a SYM background by studying their influences on the SUGRA action of the KK  modes and correlation functions, respectively.

\section{Correlation Functions From the Ansatz} \label{corr}

Due to its special role in the AdS/CFT correspondence conjectured in the previous section,
it is worthwhile to study the ansatz (\ref{ansatz}) by itself. We consider regular solutions of SUGRA following the ansatz with finite boundary values. These solutions can be used to derive  
correlation functions in SYM according to (\ref{dual}). For an arbitrary solution following the ansatz (\ref{ansatz}), the metric in Einstein frame is still the AdS metric\footnote{The stringy
correction to the supergravity action gives an $R^4$ term \cite{R4,Green} which can be written in terms of the Weyl tensor. Since the Einstein metric is still AdS$_5$, which is conformally
flat, the $R^4$ term will not introduce any correction to (\ref{F}).} and the contributions of $\phi$ and $\chi$ to the supergravity action cancel, so the right hand side of  (\ref{dual}) or (\ref{dual1}) is just $1$ (apart from the overall factor due to D-instantons). The fact that the SUGRA action
is independent of the boundary values of the fields of consideration is directly related to
the fact that the ansatz (\ref{ansatz}) preserves half of the SUSYs.

As an example, we consider the regular solutions of the ansatz (\ref{ansatz}) which have no dependence on S$^5$ and satisfy the boundary condition,
$e^{-\phi}-e^{-\phi_{\infty}} =f(y)$ at $U=\infty$.  The operator coupled to the massless dilaton field $e^{-\phi}$ is  $\O_{\phi}=\frac{1}{4}\mbox{Tr}(F^{\mu\nu}F_{\mu\nu})$ \cite{Klebanov,more}
and the operator coupled to the massless axion $\chi$ is $\O_{\chi}=\frac{1}{4}\mbox{Tr}(F^{\mu\nu}F^*_{\mu\nu})$, where $F^*_{\mu\nu}=\frac{1}{2}\eps_{\mu\nu\a\b}F^{\a\b}$. Thus we have
\be
\la e^{\int d^4 y \frac{1}{4}f(y) F^2_{\pm}(y)}\ra_{CFT} = 1,
\eq
where $F^2_{\pm}=\mbox{Tr}(F_{\pm}^{\mu\nu} F_{\pm\mu\nu})$ and   $F_{\pm\mu\nu}=F_{\mu\nu}\pm F^*_{\mu\nu}$. Whether we have the  self-adjoint part $F_+$ or
the anti-self-adjoint part $F_-$ depends on the sign in (\ref{ansatz}). It follows that, for example,
\be \label{F}
\la F_{\pm}^2(y)F_{\pm}^2(0)\ra_{CFT}=0.
\eq
To obtain the correlation function for the Minkowskian theory, we need to apply Wick rotation which maps $F_{0i}$ to $iF_{0i}$ and  $F_{ij}$ to $F_{ij}$ for $i,j=1,2,3$.
Thus
\be
F_{\pm} \;\rightarrow\; F\pm iF^*
\eq
and (\ref{F}) becomes
\bee
&\la F^2(y)\cdot F^2(0)\ra=\la FF^*(y)\cdot FF^*(0)\ra, \label{F4}\\
&\la F^2(y)\cdot FF^*(0)\ra+\la FF^*(y)\cdot F^2(0)\ra=0,
\eqq
where $F^2=\mbox{Tr}(F^{\mu\nu}F_{\mu\nu})$ and
$FF^*=\mbox{Tr}(F^{\mu\nu}F^*_{\mu\nu})$. The left hand side of (\ref{F4}) was derived in \cite{Gubser} using the AdS/CFT duality. Using their result we get
\be
\la FF^*(y)\cdot FF^*(0)\ra=\la F^2(y)\cdot F^2(0)\ra
\sim\frac{N^2}{\left| y\right|^8}.
\eq
The above derivation can be repeated using regular SUGRA solutions $\psi$ with non-trivial S$^5$ dependence which satisfy ansatz (\ref{ansatz}).  In principle we can obtain infinitely many these  identities for correlation functions of the corresponding SYM  operators $\O_{\psi}$. Being massive, these modes of higher harmonics on S$^5$ couple to  operators of higher conformal weight in the SYM \cite{chw}.

\section*{Acknowledgement}
C.S.C. thanks Ergin Sezgin and Per Sundell for discussions and the Physics Department of Cornell  University and Texas A\&M University for hospitality. P.M.H. thanks Yong-Shi Wu for discussions.  Y.Y.W. thanks Jonathan Bagger for encouragement. The work of P.M.H. was supported in part by NSF grant No. PHY-9601277.  The work of Y.Y.W. was supported in part by NSF grant No. PHY-9404057.

\end{document}